# Enhanced Light Extraction in Tunnel Junction Enabled Top Emitting UV LEDs

Yuewei Zhang,[1,a] Andrew Allerman,[2] Sriram Krishnamoorthy,[1] Fatih Akyol,[1] Michael W. Moseley,[2] Andrew Armstrong,[2] and Siddharth Rajan[1,a]

[1] Department of Electrical and Computer Engineering, The Ohio State University, Columbus, Ohio, 43210, USA

[2] Sandia National Laboratories, Albuquerque, New Mexico 87185, USA

**Abstract:** The efficiency of ultra violet LEDs is critically limited by the absorption losses in p-type and metal layers. In this work, surface roughening based light extraction structures are combined with tunneling-based top-contacts to realize highly efficient top-side light extraction efficiency in UV LEDs. Surface roughening of the top n-type AlGaN contact layer is demonstrated using self-assembled Ni nano-clusters as etch mask. The top surface roughened LEDs were found to enhance external quantum efficiency by over 40% for UV LEDs with a peak emission wavelength of 326 nm. The method described here can enable highly efficient UV LEDs without the need for complex manufacturing methods such as flip chip bonding.

III-Nitride ultra-violet light emitting diodes (UV LEDs) have attracted great research interest due to the large range of applications including water purification, air disinfection, curing, and phototherapy.[1] In the past decade, because of the improvement in substrate quality and optimization of both the active region design and fabrication, LED efficiency and power have

---

a) Authors to whom correspondence should be addressed.
   Electronic mail: zhang.3789@osu.edu, rajan@ece.osu.edu

improved significantly.[2-7] However, the efficiency of UV LEDs remains significantly lower than that of the InGaN-based visible LEDs, especially in the UV-C wavelength range.[8]

One important factor limiting UV LED performance is the low light extraction efficiency caused by light absorption in the p-type contact region and internal light reflection at the interfaces. Conventional UV LEDs use either p-GaN or p-Al$_x$Ga$_{1-x}$N/ Al$_y$Ga$_{1-y}$N superlattice layer for top contact, which induces light absorption loss or increases operating voltage. As a solution, light is most efficiently extracted from the substrate side of the devices using flip chip bonding.[2-7] To reduce internal reflection, micro-structure patterning has to be done on the back side of the devices, and device encapsulation and packaging is necessary. Despite these techniques, the achieved light extraction efficiency is only 25%[8], which is much lower than the visible LEDs (> 80%)[9]. The difference is largely attributed to the fact that the thick p-type contact layer absorbs UV light, but is transparent to visible light.

In this work, we show that nano-columns patterned on top of a tunnel junction LEDs enable highly efficient light extraction from the top of the wafer, eliminating the need for backside roughening and reflective contacts. Using the nano-column enhanced tunnel junction UV LED, we demonstrate extraction efficiency enhancement of up to 43% at 326 nm.

Tunneling based hole injection has been investigated and demonstrated in previous work for visible LEDs by both MBE and MOCVD growth techniques.[10-16] Recently, tunneling injection into a UV LED structure was demonstrated using a polarization engineered p-AlGaN/ InGaN/ n-AlGaN interband tunnel junction integrated on the thin p-AlGaN cladding layer of a UV LED structure.[17,18] The strong polarization field caused by the polarization dipole at the AlGaN/ InGaN interface enables abrupt band bending in the InGaN layer. The valence band edge in the p-AlGaN side is aligned with the conduction band edge in the n-AlGaN side within a few nanometers, forming an almost broken gap alignment with reduced interband tunneling barrier height across the thin InGaN layer.[17-19] The tunnel junction (TJ) structure enables low-resistance contact to the p-AlGaN layer with a transparent n-AlGaN top contact layer. Using this approach, photons from the active region can be extracted directly from the top surface with minimal optical loss from the thin InGaN layer.[17]

The tunneling injected UV LED structure was grown by plasma assisted molecular beam epitaxy (PAMBE) on unintentionally-doped metal-polar $Al_{0.3}Ga_{0.7}N$ template, which was grown on a sapphire substrate using metal-organic chemical vapor deposition.[20] The internal quantum efficiency is strongly affected by the high threading dislocation density of $2.8\times10^9$ $cm^{-2}$. The epitaxial structure consists of a 500 nm Si doped $Al_{0.3}Ga_{0.7}N$ bottom contact layer for electron injection, three $Al_{0.2}Ga_{0.8}N$/ $Al_{0.3}Ga_{0.7}N$ quantum wells (QWs), 50 nm p-$Al_{0.3}Ga_{0.7}N$ cladding layer, a p-AlGaN/ InGaN/ n-AlGaN tunnel junction layer, and a 300 nm n-type $Al_{0.3}Ga_{0.7}N$ top contact layer. Details of the growth conditions are reported elsewhere.[17]

Surface roughening was carried out using self-assembled Ni nano-clusters as etching mask.[21-24] The device process flow is schematically shown in Fig. 1. First, 300 nm $SiO_2$ was deposited on the as-grown sample using plasma-enhanced chemical vapor deposition (PECVD), followed by 10 nm Ni on $SiO_2$ using e-beam evaporator. The sample was annealed at 860 °C for 90 s under $N_2$ ambient to form nano-sized Ni clusters.[21-24] The self-assembled Ni nano-clusters were used as nanoscale etch masks for inductively coupled plasma reactive ion etching (ICP-RIE) of both $SiO_2$ and $Al_{0.3}Ga_{0.7}N$. $CF_4$ and $BCl_3$/ $Cl_2$ based etch chemistries were used to dry etch $SiO_2$ and $Al_{0.3}Ga_{0.7}N$, respectively. The etch depth was controlled to remove 300 nm $SiO_2$ and 150 nm $Al_{0.3}Ga_{0.7}N$, retaining 150 nm of the n-$Al_{0.3}Ga_{0.7}N$ top contact layer. The Ni nano-clusters were then etched using dilute nitric acid. The remaining $SiO_2$ nano-columns out of the device mesa were etched using buffered oxide etch (BOE) as shown in Fig. 1(e). The device mesas were then isolated using ICP-RIE etch, followed by bottom contact deposition of Ti(20 nm)/ Al(120 nm)/ Ni(30 nm)/ Au(50 nm) and subsequent annealing at 850 °C for 30 sec, and Al(20 nm)/ Ni (20 nm)/ Au(80 nm) top contact deposition. Partial metal coverage (23% electrode coverage for 50 × 50 µm$^2$ devices) for top contacts was used to minimize light absorption in the metal contact. Devices with and without surface roughening were compared on the same sample by avoiding Ni deposition in half of the sample area. The electroluminescence (EL) spectrum and emission power were measured on-wafer from the top surface of the devices at room temperature using a calibrated Ocean Optics USB 2000 spectrometer coupled with a fiber optic cable. Since no integrating sphere was used, the measurements underestimated the actual device output power.

Figure 2 shows the scanning electron microscopy (SEM) images of the sample surface after roughening. The roughening process creates randomly distributed nano-columns composed of 300 nm SiO$_2$ and 150 nm n-Al$_{0.3}$Ga$_{0.7}$N, with the interface seen clearly in the SEM images. The average diameter of the nano-columns is 300 nm, and the distance between them varies in the range of 50 nm to 100 nm. The controlled anisotropic etch profile resulted in SiO$_2$ whiskers with tapered sidewalls that increases light extraction. In addition, the higher refractive index of SiO$_2$ ($n_s$ ~ 1.5) compare to air increases the critical angle of the UV light at the Al$_{0.3}$Ga$_{0.7}$N/ SiO$_2$ interface.

The electrical characteristics were measured for 50 × 50 µm$^2$ devices with partial metal coverage. Pulsed measurements with pulse width and period of 1ms and 100 ms were performed to reduce self-heating effects. The smooth and roughened devices show similar electrical behavior, with a voltage drop of 5.15 V at 20 A/cm$^2$ (Fig. 3). The devices are more resistive than our previous report mainly due to extra voltages for current spreading in both bottom and thin top contact layers, even though the tunnel junction resistance was shown to be as low as $5.6\times10^{-4}$ $\Omega$ cm$^2$.[17] This can be mitigated by using thicker bottom/ top contact layers to reduce the spreading resistances.

On-wafer optical power measurements of the devices with/ without surface roughening were carried out at room temperature to compare the change in light extraction efficiency. As shown in Fig. 4(a), single peak emission from the active region with peak wavelength at 326 nm was achieved. Both devices showed uniform light emission (inset of Fig. 4(c)). At the same injection current of 5 mA, the rough region gives a higher emission peak due to increased light scattering at the surface. The inset to Fig. 4(a) shows the EL peak in log scale in a broad wavelength range. There is a weak peak at 510 nm for both devices, which could come from the thin InGaN layer or from a defect level in the p-AlGaN layer. The device with surface roughening reaches 1.39 mW at 20 mA under pulse mode, corresponds to a power density of 55.6 W/cm$^2$ at 800 A/cm$^2$.

As shown in Fig. 4(b) and (c), the external quantum efficiency (EQE) and wall-plug efficiency (WPE) curves indicate a light extraction enhancement of ~ 43%. The peak EQE and WPE of the roughened device are 2.21% and 1.36% under CW operation, while the values are 1.55% and 1.00% for the smooth device. This enhancement was confirmed on multiple devices on several samples (data not shown here).

While many approaches have been used to increase the light extraction efficiency of UV LEDs, all those methods have to be done on the backside of the devices after flip-chip bonding.[3,4,25] The absorbing p-type contact layer is an important reason that the light extraction efficiency of UV LEDs remains much lower than visible LEDs.[8,9] The tunneling injected UV LED structure mitigates the absorption loss issues in UV LEDs, since the InGaN interband tunneling layer used (< 4 nm) is significantly thinner than typical p-GaN capping layers (> 20 nm).[17] This makes it possible to achieve UV light extraction efficiency comparable to the visible counterparts, and reduces the complexity and cost of device fabrication. Engineering the active region design could lead to preferential surface emission instead of side-emission for deep UV LEDs emitting longer wavelengths than 240 nm.[26] In this sense, the enhanced top surface emission enabled by tunnel junction UV LED structure could be efficiently applied to UV LEDs emitting at a wide UV range.

In conclusion, we have demonstrated a 326 nm UV LED structure with tunnel injection of holes, with enhanced light extraction through surface roughening directly on the top n-AlGaN surface. Self-assembled Ni nano-clusters were used as etching mask to achieve $SiO_2$/ AlGaN nano-columns with an average diameter of ~ 300 nm on the top surface. The obtained peak EQE values from on-wafer DC measurement increases from 1.55% to 2.21% after surface roughening, resulting in a 43% increase in the light extraction efficiency. The power density reaches 55.6 $W/cm^2$ at 800 $A/cm^2$ under pulse measurement conditions. This demonstration shows that tunneling injected UV LED structures could enable improved light extraction efficiency, which is critical for high power and high efficiency UV emitters.

Acknowledgement: S.R., Y.Z., S.K. and F.A. acknowledge funding from the National Science Foundation (ECCS-1408416). Sandia National Laboratories is a multi-program laboratory managed and operated by Sandia Corporation, a wholly owned subsidiary of Lockheed Martin Corporation, for the U.S. Department of Energy's National Nuclear Security Administration under contract DE-AC04-94AL85000.

Figure Captions:

Fig. 1 Schematic device processing flow. (a) Growth stack of the tunnel junction UV LED structure. (b) Deposition of 300 nm $SiO_2$ and 10 nm Ni. (c) Annealing to form Ni nano-clusters. (d) ICP-RIE etch of $SiO_2$ and n-AlGaN top layer to form nano-columns, and Ni removal using Nitric acid. (e) Define mesa region and wet etch of $SiO_2$. (PR represents photoresist) (f) Device finalization.

Fig. 2 Scanning electron microscopy (SEM) image of the roughened top surface after 300 nm $SiO_2$/ 150 nm $Al_{0.3}Ga_{0.7}N$ selective etch and Ni removal using Nitric acid.

Fig. 3 Electrical characteristics for smooth and roughened tunnel junction UV LED devices (50 × 50 µm$^2$) under both CW and pulse (pulse width and period of 1 ms and 100 ms) measurements.

Fig. 4 Optical characteristics for smooth and roughened tunnel junction UV LED devices (50 × 50 µm$^2$) under CW and pulse modes. (a) Electroluminescence (EL) with CW current injection of 5 mA at room temperature, single peak emission at 326 nm is shown. The inset shows the EL spectrums in log scale, a small long wavelength peak is shown at 510 nm. (b) Output power, (c) EQE and (d) WPE of the devices. The powers were measured on wafer without integrating sphere from the top surface. The inset to (c) is a microscope image of a 50 × 50 µm$^2$ TJ-UV LED device under 10 mA injection current (dark region outlines the metal contact).

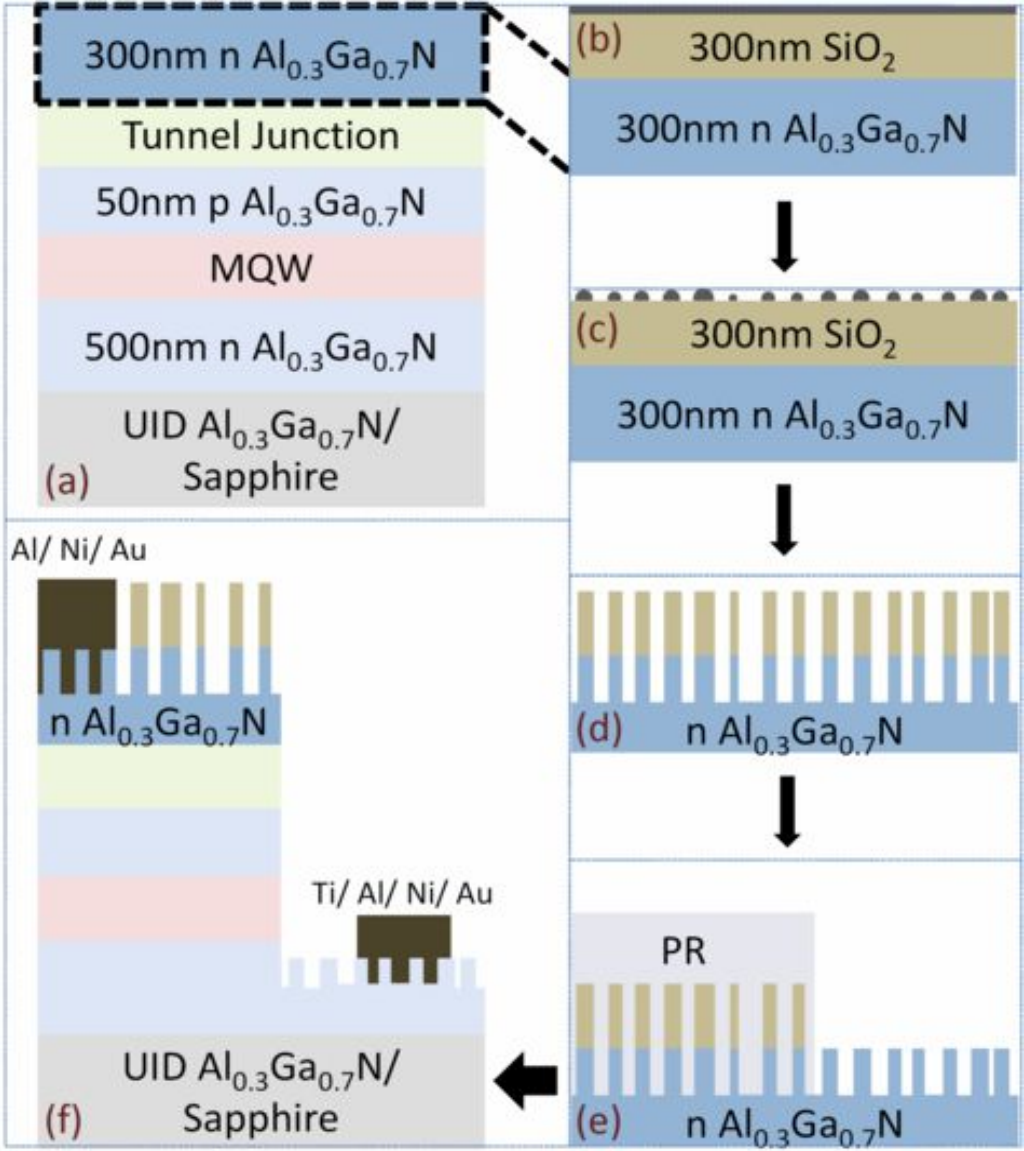

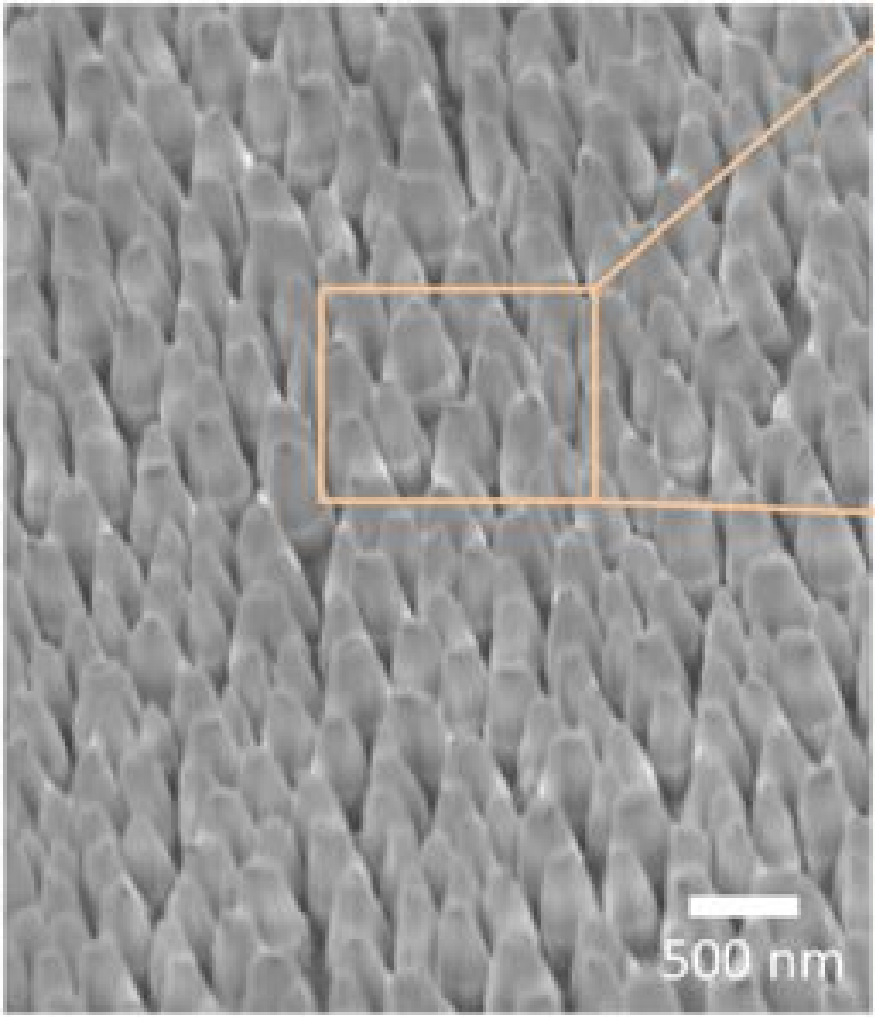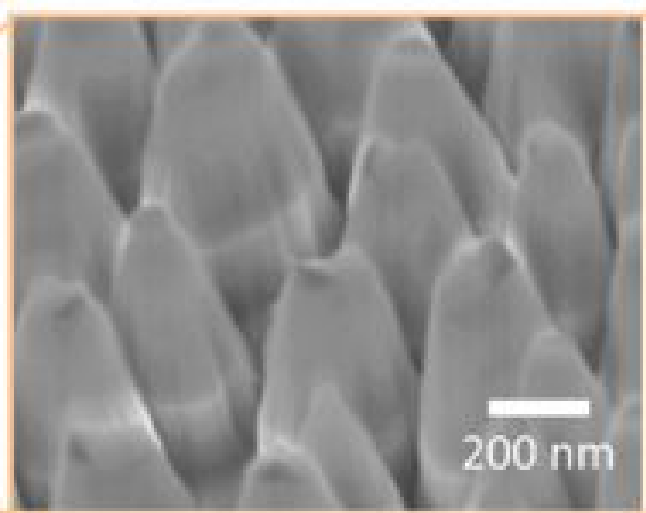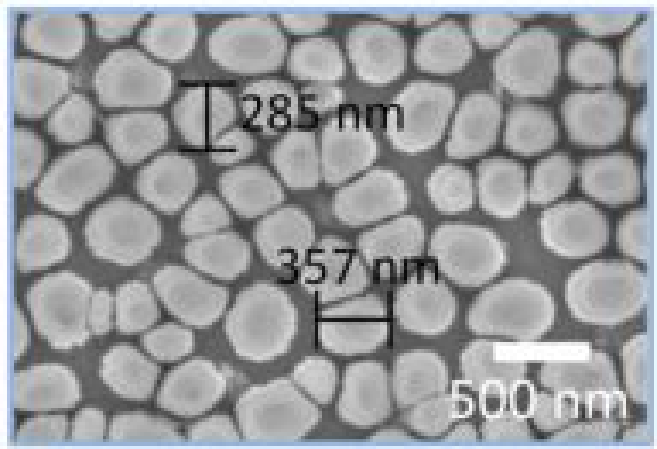

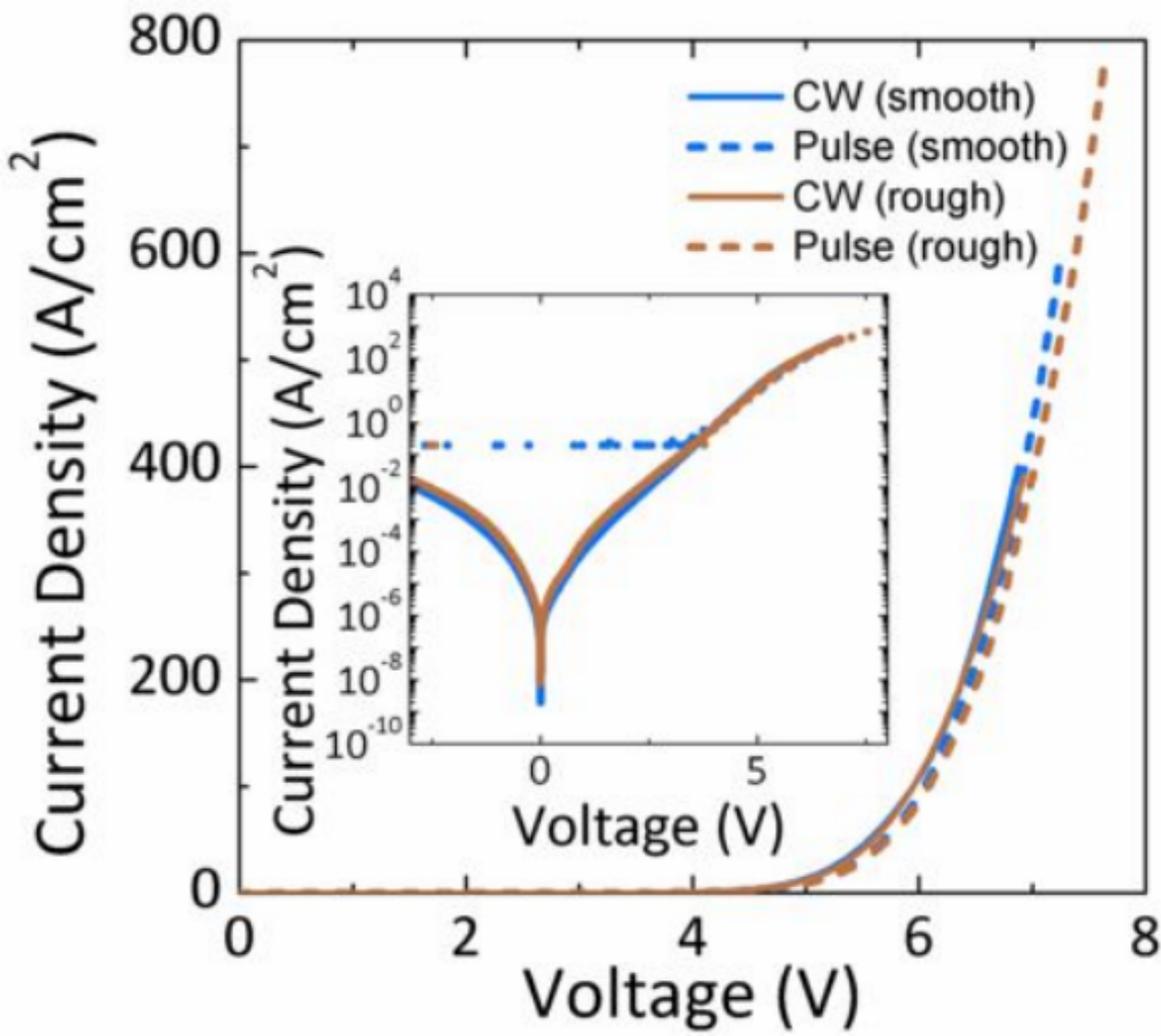

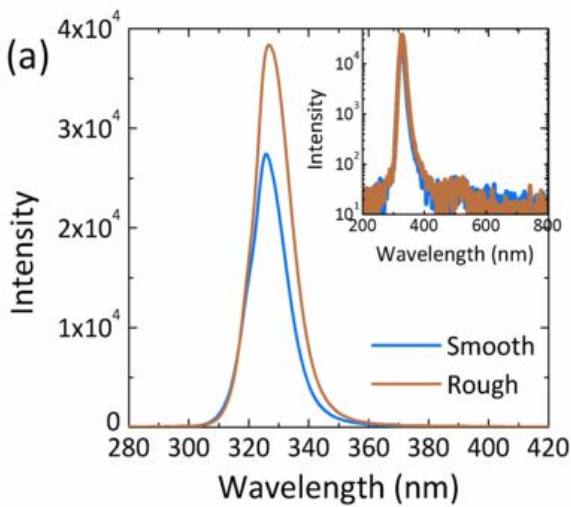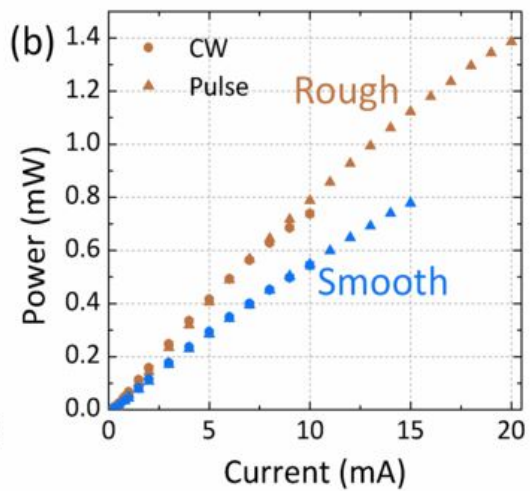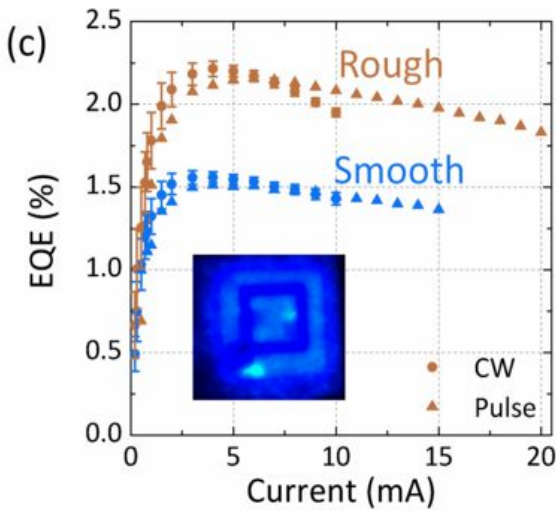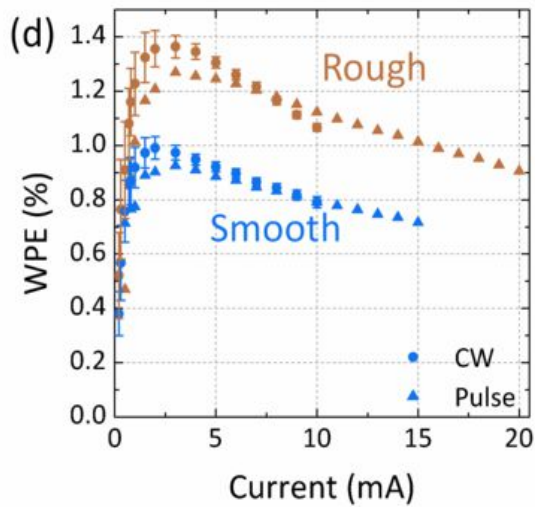